\documentclass[aps,prd,onecolumn,superscriptaddress,showpacs,bibnotes,preprintnumbers,amsmath,amssymb]{revtex4}
\usepackage{graphicx}
\usepackage{dcolumn}
\usepackage{bm}
\usepackage{epstopdf}
\usepackage{mathtools}
\usepackage{natbib}
\usepackage{captcont}
\usepackage{booktabs}
\usepackage{threeparttable}

\begin{document}
\title{Limits on the Dark Matter from AMS-02 antiproton and positron fraction data}
\affiliation{School of Physics, Nanjing University, Nanjing, 210093, China}
\affiliation{Key Laboratory of Dark Matter and Space Astronomy, Purple Mountain Observatory, Chinese Academy of Sciences, Nanjing 210008, China }
\author{Bo-Qiang Lu}
\affiliation{School of Physics, Nanjing University, Nanjing, 210093, China}
\affiliation{Key Laboratory of Dark Matter and Space Astronomy, Purple Mountain Observatory, Chinese Academy of Sciences, Nanjing 210008, China }
\author{Hong-Shi Zong}
\affiliation{School of Physics, Nanjing University, Nanjing, 210093, China}

\begin{abstract}
We derive limits on the dark matter annihilation cross section and lifetime using measurements of the AMS-02 antiproton ratio and positron fraction data. In deriving the limits, we consider the scenario of secondary particles accelerated in supernova remnants (SNRs) which has been argued to be able to reasonably account for the AMS-02 high energy positron/antiproton fraction data. We parameterize the contribution of secondary particles accelerated in SNRs and then fit the observational data within the conventional cosmic ray propagation model by adopting the GALPROP code.
We use the likelihood ratio test to determine the 95$\%$ confidence level upper limits of the possible dark matter (DM) contribution to the antiproton/positron fractions measured by AMS-02. Our limits are stronger than that set by
the Fermi-LAT gamma-ray Pass 8 data of the dwarf spheroidal satellite galaxies. 
We also show that the solar modulation (cosmic ray propagation) parameters can play a non-negligible role in modifying the constraints on the dark matter annihilation cross section and lifetime for $m_\chi<100$ GeV ($m_\chi>100$ GeV), where $m_\chi$ is the rest mass of the dark matter particles.
Using this results, we also put limits on the effective field theory of dark matter.
\end{abstract}
\maketitle

\section{Introduction}
Eighty years after its discovery \cite{Ade2014apj, D.Hooper2005, BBN, rotcur, gravlen} the nature of dark matter (DM) remains to be mysterious. Among various hypothetical particles, the so-called weakly interacting massive particle (WIMP) is the leading
candidate \cite{D.Hooper2005}. It is also widely believed that WIMPs could annihilate with each other and then generate (or alternatively decay into) stable particles, such as high energy $\gamma-$rays and pairs of electrons/positrons, protons/antiprotons, and neutrinos/antineutrinos \cite{D.Hooper2005, Griest2000, Rouven}. Such particles propagate into the Galaxy and then become part of the cosmic rays. The accurate measurements of cosmic rays hence in turn provides the people a valuable chance to study the dark matter particle indirectly. The indirect detection of dark matter particles with space-based cosmic ray detectors has been a quickly evolving field since 2008 \cite{ATIC, PAMELA1, PAMELA2, AMS021, AMS022, AMS023, AMS024}. The most-extensively discussed signature is the high energy spectra of cosmic ray electrons and positrons (equally, the positron to electron ratio data) that are well in excess of the prediction of the conventional cosmic ray propagation model, i.e., the so-called electron/positron excesses \cite{yizhongfan2010}. Before 2015, due to the lack of evidence for an antiproton excess \cite{PAMELAantp1}, the electron/positron excesses have been widely attributed to the leptonic dark matter annihilation/decay \cite{ackermann2012, fengl2013, yuanq2014, khlopov2014}. Nevertheless, the astrophysical origins, such as the electron/positron pairs from pulsars \cite{Cholis} or the secondary particles accelerated in the SNRs \cite{Blasiprl1, Blasiprl2, Philipp2009prl, Philipp2014prd}, can also reasonably account for the data. For example, Di Mauro et al. \cite{Mauro2014} has shown that the electron/positron data can be satisfactorily modeled through the sum of an average primary electron flux from distant sources and the fluxes from the local supernova remnants in the Green catalog and the secondary electron and positron fluxes originate from interactions on the interstellar medium of primary cosmic rays. Such facts suggest that in comparison to identifying dark matter signal it may be more reasonable to use the data to set limits on the
dark matter annihilation cross section and decay lifetime in 
given dark matter scenarios. In the PAMELA era, in view of the lack of distinct spectral structures that are predicted to arise in dark matter annihilations/decays into electrons/positrons in the data, some researchers constrained the physical parameters of these exotic particles by assuming that the positron fraction excess arises from a group of pulsars \cite{fengl2013}. With the first AMS-02 result (i.e., the positron fraction up to $\sim 350$ GeV \cite{AMS021}), an improved approach yielded stringent limits on dark matter annihilating or decaying to leptonic final states \cite{Bergstrom2013prl}. Later,  the positron flux data or alternatively the electron flux data has been adopted to set limits on the dark matter annihilation/decay channels \cite{Alejandro2013, lu2015}. Recently, the AMS-02 antiproton-to-proton ratio data has been announced in a dedicated conference \cite{AMS02Aprtalk} and the high energy part seems to be in excess of the regular prediction of conventional cosmic ray propagation model (while in Ref.\cite{Giesen2015, Kappl2015} they argue that the background (BKG) is enough to explain AMS-02 antiproton data if uncertainty in propagation model, solar modulation and nucleon collision process are taken into account). In this work we take both the antiproton ratio data and the positron fraction data to place the limits on the physical parameters of dark matter particles.
The pulsar model may account for the positron fraction excess but likely not for the antiproton data. In this work we consider the scenario of secondary particles accelerated in SNRs \cite{Blasiprl1, Blasiprl2} which may explain both the AMS-02 positron fraction and antiproton ratio data. Following the phenomenological AMS parametrization approach we parameterize the contribution of the SNR with a simple function and calculate the background (BKG) ratio of positron and antiproton with GALPROP\cite{galprop}, then we add the SNR component to the BKG component as the total ratio at top of the atmosphere (TOA).  Using this result we place limits on the DM parameters. We find that our limits are stronger than the limits given by Ackermann et al. \cite{Ackermann} which derived from the Fermi-LAT gamma-ray Pass 8 data on the dwarf spheroidal satellite galaxies.
Further more, we use this results to put constrains the effective field theory which are mostly discussed at direct detection and Large Hadron Collider (LHC).

This work is arranged as the following. In Section II we briefly introduce the model of secondary particle acceleration in SNR, some details of BKG CR propagation model. In Section III we present our limits on the dark matter parameters and study the uncertainties caused by the propagation parameters, solar modulation parameters, and the dark matter distribution profile models. And in Section IV we put limits on the effective field theory by using the results we obtain in Section III. We summarize our results in Section V.

\section{The astrophysical origin model of AMS-02 antiproton ratio and positron fraction data}

\subsection{Secondary particle acceleration in SNR}

In this work, we consider the acceleration of secondary CRs in the SNR diffusive shock. The evolution of the gyro-phase and pitch-angle averaged phase space density $f_{\rm i} \equiv f_{\rm i}(x, p)$ of species $i$ is governed by the transport equation \cite{Blasiprl1, Blasiprl2, Philipp2009prl, Philipp2014prd}

\begin{eqnarray}
\frac{\partial f_{\rm i}}{\partial t}=-u\frac{\partial f_{\rm i}}{\partial x}+\frac{\partial }{\partial x}D_{\rm i}\frac{\partial f_{\rm i}}{\partial x}-\frac{p}{3}\frac{du}{dx}\frac{\partial f_{\rm i}}{\partial p}+Q_{\rm i}(x, p) ,
\end{eqnarray}
and the the solution of this equation is found to be

\begin{eqnarray}
f_{\rm i}(x=0, p)=\gamma (\frac{1}{\xi }+r^{2})\int_{0}^{\rm p}\frac{d{p}'}{{p}'}(\frac{{p}'}{p})^{\gamma }\frac{D_{\rm i}({p}')}{u_{-}^{2}}Q_{\rm i}({p}'),
\end{eqnarray}
where $-$ represent upstream and $+$ represent downstream, the slope $\gamma =3u_{-}/(u_{-}-u_{+})=3r/(r-1)$, $u$ is the velocity of fluid, $r=u_{-}/u_{+}$ is the compression factor. For a strong shock $r\rightarrow 4$ and $\gamma \rightarrow  4$.  The factor $\xi $ represents the mean fraction of the energy of an accelerated proton carried away by a secondary particle in each scattering. $D_{\rm i}(p)\propto p^{\alpha }$ is the diffusion coefficient of the shock \cite{Blasiprl1}. The production rate at a position $x$ around the shock is

\begin{eqnarray}
Q_{\rm i}(x, E)=\sum _{\rm j}\int d\sigma _{\rm ji}({E}', E) cN_{\rm j}(x, {E}')n_{\rm gas}(x),
\end{eqnarray}
where $c$ is the speed of light, $\sigma _{\rm ji}({E}', E)$ is the cross section for a primary specie $j$ of energy ${E}'$ to produce a secondary particle {\it i} of energy $E$. The source spectrum $N_{\rm j}=4\pi p^{2}f_{\rm j}(p)u_{+}\tau  _{SN}$ and $f_{\rm j}\sim p^{-\gamma }$ then $N_{\rm j}\sim p^{-\gamma +2}$. $n_{\rm gas}$ is the gas density in the shock region. Then one can find that $f_{\rm i}(x=0, p)\sim p^{-\gamma +\alpha}$ and $\alpha > 0$ is the slope of the diffusion coefficient (in the following $\alpha $ is taken as 1 for a Bohm-like diffusion coefficient), this result indicates that he equilibrium spectrum of the particles that take part in the acceleration is flatter than the injection spectrum of secondary particles \cite{Blasiprl1, Philipp2014prd}.
As presented in \cite{Blasiprl2} the secondary-to-primary ratio (such as $\bar{p}/p$) which contributed from accelerating in the SNR can be easily derived with above results

\begin{eqnarray}
\mathcal{R}_{\rm SNR}(E)\simeq cn_{\rm gas}[\mathcal{A}(E)+\mathcal{B}(E)],
\end{eqnarray}
where
\begin{eqnarray}
\mathcal{A}(E)=\gamma (1/\xi +r^{2})\int_{m}^{E}dyy^{\gamma -3}\frac{D_{-}(y)}{u_{-}^{2}}\int _{\rm y}^{E_{\rm max}}dzz^{2-\gamma }\sigma _{\rm ji}(z, y),
\end{eqnarray}
and
\begin{eqnarray}
\mathcal{B}(E)=\frac{\tau _{\rm SN}r}{2E^{2-\gamma }}\int_{E}^{E_{\rm max}}dzz^{2-\gamma }\sigma _{\rm ji}(z, E).
\end{eqnarray}
One can easily find that $\mathcal{B}(E)$ term is nearly a small constant and has been neglected in this work, and $\mathcal{A}(E)\propto E^{-\kappa  +2}$ where $\kappa $ represent a factor of nucleon collision which means that the slope of $\mathcal{A}(E)$ term dose not depend on the accelerated process but only decided by the nucleon collision. For the case of antiproton $\mathcal{A}(E)\propto E^{1.55}$ in \cite{Blasiprl2} indicated that $\kappa\sim 0.5$ and $\kappa\sim 0.4-0.6$ when 40$\%$ uncertainty in nucleon collision process is taken into account. 
We note here that the Eq.4 is also suitable for the case of positron fraction, the reason is that since positrons and electrons suffered nearly the same energy loss after releasing into the interstellar space, so the energy losses of positrons and electrons are cancelled out each other in the positron fraction. This conclusion may be seen from the BKG component of positron fraction in the right panel of FIG.1.
Spallation and decay are taken into account in \cite{Philipp2009prl, Philipp2014prd} (add a term $-\Gamma _{\rm i}f_{\rm i}$ on the right of Eq.1) which lead to a suppression of the secondary contribution at very high energies. 
In this work we take a form of 
\[\mathcal{R}_{\rm SNR}(E)=\mathcal{N}_{\rm SNR}(E/1GeV)^{-\kappa+2}{\rm exp}(-E/E_{\rm c})\] 
as the contribution of secondary particles accelerated in SNR, then the observed ratio of secondary-to-primary is $\mathcal{R}_{\rm OBS}=\mathcal{R}_{\rm BKG}+\mathcal{R}_{\rm SNR}$.

\subsection{Background cosmic ray propagation model}
Primary CRs are released into the interstellar space after accelerated in SNR sources and the secondaries CRs are produced by the interaction of primary with the ISM, the propagation of CRs in the galaxy was described by the transport equation \cite{strong1998, strong2007}

\begin{eqnarray}
\frac{\partial \psi }{\partial t}=\triangledown \cdot (D_{\rm xx}\triangledown\psi -V_{\rm c}\psi )+\frac{\partial }{\partial p}p^{2}D_{\rm pp}\frac{\partial }{\partial p}\frac{1}{p^{2}}\psi -\frac{\partial }{\partial p}\begin{bmatrix}
 \dot{p}\psi -\frac{p}{3}(\triangledown\cdot V_{\rm c}\psi)
\end{bmatrix}-\frac{\psi }{\tau _{\rm f}}-\frac{\psi }{\tau _{\rm r}}+Q(x,p),
\end{eqnarray}
where $\psi (\overrightarrow{r}, p, t)$ is the CR density per unit of total particle momentum $p$ at position $\overrightarrow{r}$, $D_{\rm xx}$ is the spatial diffusion coefficient that can be parameterized as $D_{\rm xx}=D_{0}\beta (R/R_{0})^{\delta }$, where $\beta=v/c$ and $R=pc/Ze$ is the particle rigidity,
We use the default setting of the propagation parameters in the GALPROP $D_{0}=5.3 \times 10^{28}$ ${\rm (cm^{2}s^{-1})}$, $R_{0}=4.0$ ${\rm GeV}$ and  $\delta=0.33$, such values can fit the observational $\rm B/C$, $^{10}\textrm{Be}/^{9}\textrm{Be}$ well. 
However, there exist degeneracies of different set of propagation parameters which cannot be distinguished with present experiments,
so we also consider the effect of the propagation parameters on the exclusion line in the following.
$D_{\rm pp}$ is diffusion coefficient in the momentum space and is related to $D_{\rm xx}$ by

\begin{eqnarray}
D_{\rm pp}D_{\rm xx}=\frac{4p^{2}v_{A}^{2}}{3\delta (4-\delta ^{2})(4-\delta )\omega },
\end{eqnarray}
where $\omega$ characterizes the level of turbulence and is taken as $1$, $v_{A}$ is Alfv$\acute{\rm e}$n speed which is set at 33.5 $\rm kms^{-1}$ in the DR scheme.
 $V_{\rm c}$ is the convection velocity that is assumed to increase linearly with distance from the plane \cite{strong1998, Zirakashvili1996}, $\triangledown \cdot V_{\rm c}$ represents adiabatic momentum gain or loss in the non-uniform flow of gas with a frozen-in magnetic field whose inhomogeneities scatter the CRs \cite{strong2007}, and $\dot{p}=dp/dt$ is the momentum gain or loss rate, $\tau _{\rm r}$ is the characteristic time scale for radioactive decay and $\tau _{\rm f}$ is the characteristic time scales for loss by fragmentation. $Q(x,p)$ is the source term including primary, spallation and decay contributions. The distribution of CR sources is taken as,

\begin{eqnarray}
q(r, z)=q_{0}(\frac{r}{r_{\odot }})^{\eta }{\rm exp}(-\xi \frac{r-r_{\odot }}{r_{\odot }}-\frac{\left | z \right |}{z_{\rm s}}).
\end{eqnarray}
where $q_{0}$ is a normalization constant, $r_{\odot }$ = 8.5 kpc is the solar position in the galaxy, $z_{\rm s}=0.2$ kpc, $\eta = 1.69$ and $\xi =3.33$, $r$ is the galaxy radius and a cut-off had been used in the source distribution at $r = 20$ kpc since it is unlikely that significant sources are present at such large radii, $z$ is the column height of the galaxy and its maximum value is set to be $z_{h}$ = 4 kpc.
The diffusion re-acceleration (DR) propagation model has been adopted in this work.

We point out that the same as the SNR component the BKG component can also be got by using phenomenological AMS parametrization approach \cite{Bergstrom2013prl}, or on the other hand the propagation of the SNR component (secondary particle accelerated in SNR as a source) can also be calculated using GLPROP (and a nice fit results of positron fraction and antproton ratio can be found in Ref.\cite{Philipp2014prd}). The former is a fully parametrization method while the latter is a fully physical method, the limits do not change significantly between this two methdos \cite{Bergstrom2013prl}, because the ratio (or flux) at TOA is the same for the two methods. The aim of our method present in this work is to get the right ratio or fraction at TOA, while providing the physical processes in some details at the same time. We also note that there are also some other astrophysical models \cite{kohri2015} which may account for the AMS-02 data well, but we do not discuss them in details here.

\section{Limits On The Dark Matter Parameters}
We consider DM annihilations or decays in the following channels,
\[\chi \bar{\chi }\rightarrow b\bar{b},u\bar{u},W^{+}W^{-},\mu ^{+}\mu ^{-},\tau ^{+}\tau ^{-} {\rm (annihilation)}\]
\[\chi \rightarrow b\bar{b},u\bar{u},W^{+}W^{-},\mu ^{+}\mu ^{-},\tau ^{+}\tau ^{-} {\rm (decay)}\]	
each with 100$\%$ branching ratio. The annihilations or decays of dark matter particles in the Milky Way dark matter halo at the position $\overrightarrow{r}$ with respect to the Galactic center produce a primary flux with a rate (per unit energy and unit volume) that is given by \cite{DMsource}
\begin{eqnarray}
Q_{\rm anni}(\overrightarrow{r}, E)&=&\frac{\left \langle \sigma v \right \rangle}{2m_{\chi }^{2}}\frac{dN}{dE}\times \rho ^{2}_{\chi }(\overrightarrow{r}) , \\
Q_{\rm decay}(\overrightarrow{r}, E)&=&\frac{1}{\tau m_{\chi }}\frac{dN}{dE}\times \rho_{\chi }(\overrightarrow{r}),
\end{eqnarray}
where $m_{\chi}$ is the DM mass, $\tau$ is the DM particle lifetime, $\left \langle \sigma v \right \rangle$ is the DM velocity-weighted annihilation cross section, ${dN}/{dE}$ is the energy spectrum of SM particles produced in the annihilation or decay of DM particles which we simulate using the event generator PYTHIA package \cite{PYTHIA}, and $\rho_{\chi }$ is the density of dark matter particles in the Milky Way halo.
We note that the predictions of cosmic-ray fluxes originated from DM annihilations or decays crucially depends on the distribution of DM
in the galactic halo, however this astrophysical uncertainty is irreducible presently. In this work the profile is adopted to be the Navarro-Frenk-White (NFW) distribution \cite{NFW}, the Einasto and Isothermal distribution are also in consideration as a comparision.
\begin{eqnarray}
\rho (r)=\frac{\rho _{s}}{(r/r_{\rm s})(1+r/r_{\rm s})^{2}},
\end{eqnarray}
where $r_{\rm s}=20$ kpc and $\rho_{s}=0.26$ GeV $\rm cm^{-3}$. Such a value of $\rho_{s}$ corresponds to a local DM energy density of 0.3 GeV $\rm cm^{-3}$ \cite{yuanq2014}. We note that currently the more likely value is 0.43 GeV~cm$^{-3}$. Such a correction of course is minor but the limits would be tighter by a factor of 2 (annihilation model) or 1.4 (decay model).

The statistical method of likelihood ratio test developed in \cite{likelyh} is adopted to put limits on a possible DM contribution to the data measured by AMS-02. 
the likelihood function $\mathcal{L}(\vec{\theta })$ is taken the form as,
\begin{eqnarray}
\mathcal{L}(\vec{\theta })={\rm exp}(-\chi ^{2}(\vec{\theta })/2),
\end{eqnarray}
where
$\vec{\theta }=\begin{Bmatrix}
\theta ^{1},\theta ^{2},\cdots ,\theta ^{n}
\end{Bmatrix}$
is the parameters of the model, and the $\chi ^{2}(\vec{\theta })$ function is
\begin{eqnarray}
\chi ^{2}(\vec {\theta })=\sum _{\rm i}^{\rm m}\frac{(\lambda _{\rm i}^{\rm exp}-\lambda _{\rm i}^{\rm the})^{2}}{\sigma _{\rm i}^{2}}.
\end{eqnarray}
where {\it m} is the number of data, $\lambda _{\rm i}^{\rm exp}$ is the measured value and $\lambda _{\rm i}^{\rm the}$ is the theory value for a certain model and the $\sigma _{\rm i}$ is the known deviation of the measurement.
The fit results of $\chi ^{2}/d.o.f$ for antiproton and positron are $22/24$ and $43/58$ which means a quite well fit results (see FIG.1).
Upper limits at the $95\%$ $\rm C.L.$ on the DM annihilation or decay rate are derived by increasing the signal normalization from its best-fit value of astrophysic source model we discussed above until $\chi ^{2}$ changes by 2.71 i.e.
\begin{eqnarray}
\chi ^{2}_{\rm DM}=\chi ^{2}+2.71.
\end{eqnarray}
Following this procedure, the positron fraction is used to calculate the constraints on the annihilation cross section and lifetime for the final states $\mu ^{+}\mu ^{-}$ and $\tau ^{+}\tau ^{-}$ while antiproton ratio is used to calculate the constraints on the annihilation cross section and lifetime for the final states $b\bar{b}$, $u\bar{u}$ and $W^{+}W^{-}$, the results are presented in FIG.2. We also compaire our results with the limits given by Ackermann et al. \cite{Ackermann} which derived from the Fermi-LAT gamma-ray Pass 8 data on the dwarf spheroidal satellite galaxies in the left panel of FIG.3, we find that our limits are stronger than theirs both for the $\tau ^{+}\tau ^{-}$ and $b\bar{b}$ final states.

In the right panel of FIG.3 we study the the effect of the solar modulation on our results.
After the galactic CRs entering into the solar system due to the diffusion, these particles are suffered to convection, particle drift and adiabatic energy loss in the
interplanetary magnetic field carried out by the solar wind. Such an effect is the so-called solar modulation which depends, via drifts in the large scale gradients of the solar magnetic field (SMF), on the particle charge including its sign. As a result, the solar modulation depends on the polarity of the SMF, which changes periodically every $\sim 11$ years \cite{Clem1996}. Recently the stochastic method is used to solve the four-dimensional Parker (1965) transport equation which describes the transport of charged particles in the solar system, such progresses are remarkable, however there are still some uncertainties in this theory and the code is time-consuming. So in this work we only use the force field approximation since it work well above about 0.5 GeV \cite{fisk1973}. We  consider the solar modulation uncertainty $\Delta \phi \simeq 200$ MV around the best-fit value $\phi =845$ MV, as showed in the right panel of FIG.3, the uncertainty has a utmost value about 16$\%$ at 10 GeV, then decline to zero at $\sim 300$ GeV.

In FIG.4 we study the the effect of the propagation parameters of $z_{\rm h}$, $D_{0}$ and $\delta$ on our results (Just the $b\bar{b}$ final state is considered because for the final states of $\mu ^{+}\mu ^{-}$ and $\tau ^{+}\tau ^{-}$ the results are found to be insensitive of the propagation parameters \cite{Alejandro2013}). Specifically, each time we change one parameter and fix the others to be fiducial values mentioned above. 
We find that the exclusion line alter slightly with the column height of the galaxy $z_{\rm h}$, so it may only contribute about 2$\%$ uncertainty of our results if the uncertainty $\Delta z_{\rm h}\simeq 0.5$ kpc is taken into account. 
The situation changed significantly for the diffusion parameters $D_{0}$ and $\delta$ since 
the diffusion process dominates the propagation of antiprotons in the Galaxy. For example, with an uncertainty $\Delta D_{0}\simeq 1.0\times 10^{28}$ $(\rm {cm^{2}s^{-1}})$ in $D_{0}$, our limits changes $\sim 2\%$ for $m_{\chi}\leq 30$ GeV but $\sim 16\%$ for $m_\chi \sim 30-1000$ GeV then declining to $\sim$ 10$\%$ above 1000 GeV. In the case of $\delta$ if one consider the uncertainty $\Delta \delta\simeq 0.1$ in $\delta$ then it contributes about 4$\%$ uncertainty for $m_\chi<100$ GeV, while above 100 GeV the uncertainty raise to about 14$\%$. In the right bottom panel of FIG.4 we study the the effect of DM distribution profile on the limits, we can find that it contribute about 20$\%$ uncertainty in the whole DM mass range if we consider the NFW profile as the standard DM distribution profile.
So the propagation parameters contribute most uncertainty at large DM mass (above $\sim 100$ GeV) while the most uncertainty at low DM mass is contributed from solar modulation, since the diffusion dominants the propagation of CRs at high energy while the solar modulation effect the CRs mostly at lower energy.
As a result, the uncertainty of the limits on the DM parameters is about $(20-30)\%$ in the whole DM mass range if we take into account the contributions of propagation parameters and solar modulation, this value raises to $(40-50)\%$ if the contributions of DM distribution profile has been taken into account.

The other forms of uncertainty may contribute from the energy spectrum $dN/dE$, it may be different if $dN/dE$ is generated from the PPPC 4 DM ID package \cite{pppc4}. The degeneracy between diffusion re-acceleration (DR) and diffusion convection (DC) propagation model may also contribute uncertainty to the limits results, but this maybe tiny, specifically,for the final state $b\bar{b}$, $u\bar{u}$ and $W^{+}W^{-}$ the diffusion dominants the propagation of antiproton , and for the final states $\mu ^{+}\mu ^{-}$ and $\tau ^{+}\tau ^{-}$ the difference of limit results between DR and DC maybe little especially at $m_\chi>200$ GeV \cite{lu2015}.

\section{Contrains on the effective field theory}
\subsection{Dark matter annihilation}
In the following, we use our results to put limits on the parameters of effective field theory (EFT). We assume the DM as a Dirac fermion (we note that in the above results we have assumed a self-conjugate DM particle, the limits will improve by a factor 2 for the Dirac fermion case). We also assume that the WIMPs is a singlet under the SM gauge groups, thus possesses no couplings to the electroweak gauge bosons at tree-level \cite{goodman2010}. The WIMPs may interact with the SM particles through a dark gauge sector, this symmetry is spontaneous breaking at low energy and leading to a supression of the interacton between WIMPs and SM particles. The EFT can approximatively describe such interaction by using higher-dimensional operators, and this method is model independent. But we should borne in mind that this method will be broken down when the typical reaction energy is much higher than the mediator mass. In this work we study the following EFT operators
\begin{eqnarray*}
\mathcal{O}_{1}&=&\frac{m_{\rm f}}{\Lambda ^{3}}\bar{\chi }\chi \bar{f}f\\
\mathcal{O}_{2}&=&\frac{m_{\rm f}}{\Lambda ^{3}}\bar{\chi }\gamma ^{5}\chi \bar{f}\gamma ^{5}f\\
\mathcal{O}_{3}&=&\frac{1}{\Lambda ^{2}}\bar{\chi }\gamma ^{\mu }\chi \bar{f}\gamma ^{\mu }f\\
\mathcal{O}_{4}&=&\frac{1}{\Lambda ^{2}}\bar{\chi }\gamma ^{\mu } \gamma ^{5}\chi \bar{f}\gamma ^{\mu }\gamma ^{5}f,
\end{eqnarray*}
where $f$ is a SM fermion and $m_{\rm f}$ is the mass, $\Lambda=\frac{M}{g_{\chi }g_{\rm f}}$, $M$ is the mass of the exchanged particle, $g_{\chi}$ and $g_{\rm f}$ are the couplings. Then the annihilation cross section of the operators is given by \cite{yang2010}
\begin{eqnarray*}
\left \langle \sigma _{1}v \right \rangle&=&\frac{3m_{\rm f}^{2}}{8\pi \Lambda^{6}}\sqrt{1-\frac{m_{\rm f}^{2}}{m_{\chi }^{2}}}(m_{\chi}^{2}-m_{\rm f}^{2})\left \langle v^{2} \right \rangle\\
\left \langle \sigma _{2}v \right \rangle&=&\frac{3m_{\rm f}^{2}}{16\pi \Lambda^{6}}\sqrt{1-\frac{m_{\rm f}^{2}}{m_{\chi }^{2}}}m_{\chi}^{2}\left ( 8+\frac{2m_{\chi}^{2}-m_{\rm f}^{2}}{m_{\chi}^{2}-m_{\rm f}^{2}}\left \langle v^{2} \right \rangle \right )\\
\left \langle \sigma _{3}v \right \rangle&=&\frac{1}{16\pi \Lambda^{4}}\sqrt{1-\frac{m_{\rm f}^{2}}{m_{\chi }^{2}}}\left ( 24(2m_{\chi}^{2}+m_{\rm f}^{2})+\frac{8m_{\chi}^{4}-4m_{\chi}^{2}m_{\rm f}^{2}+5m_{\rm f}^{4}}{m_{\chi}^{2}-m_{\rm f}^{2}}\left \langle v^{2} \right \rangle \right )\\
\left \langle \sigma _{4}v \right \rangle&=&\frac{1}{16\pi \Lambda^{4}}\sqrt{1-\frac{m_{\rm f}^{2}}{m_{\chi }^{2}}}\left ( 24m_{\rm f}^{2}+\frac{8m_{\chi}^{4}-22m_{\chi}^{2}m_{\rm f}^{2}+17m_{\rm f}^{4}}{m_{\chi}^{2}-m_{\rm f}^{2}}\left \langle v^{2} \right \rangle \right ),
\end{eqnarray*}
where $v$ is WIMPs relative velocity in unit c. Specificly, in the early Universe $\left \langle v^{2} \right \rangle\simeq 0.3$, while today $\left \langle v^{2} \right \rangle\simeq 10^{-6}$ (in the following, we use warm DM represents for the former case and cold DM for the latter case).
We calculate the limits on $\Lambda$ for final state of b and u quarks, $\mu $ and $\tau$ leptons, and for each final state we consider the case of warm DM and cold DM respectively. The corresponding results are showed in FIG.6. We can find that the limits of $\mathcal{O}_{2}$ operator and $\mathcal{O}_{3}$ operator are not sensitive to the relative velocity $v$.
\subsection{Dark matter decay}
As showed in right panel of FIG.2, the bound of DM lifetime is $\tau \gtrsim 10^{28}\rm s$, this indicates that the DM is stable. We can speculate that the decay of WIMPs are suppress by a very large mass scale such as Planck scale $M_{\rm pl}$.
As pointed out in \cite{yann2015} the global symmetries are generically violated at the Planck scale, to describe DM decay they proposed some dimension-five effective   
operators which violate global symmetries. By requiring the couplings $\lambda \sim \mathcal{O}(1)$, they rule out a rather large DM mass range, including the classic WIMP mass range around the electroweak scale.
We also use our results to put limits on the coupling of the operators $\rm O9$ and $\rm O15$ (see left panel of FIG.6), our results are similar to \cite{yann2015}. 

A main characteristic of the decay operators proposed in \cite{yann2015} is the decay final states of a DM only contain SM particles, in the following we consider a DM may decay to another DM and SM particles at the same time, but without taking into account the global symmetries. We consider a $\rm V-A$ effective interaction
\begin{eqnarray}
\mathcal{H}_{\rm VA}=\frac{\lambda^{2}}{M_{\rm pl}^{2}}\bar{\varphi _{1}}\gamma _{\mu }(1-\gamma _{5})\chi _{1}\bar{\chi _{2}}\gamma ^{\mu }(1-\gamma _{5})\varphi _{2},
\end{eqnarray}
where $\varphi $ represents a SM particle. It decribes the decay process $\chi_{1}\rightarrow \chi_{2}\varphi_{1}\varphi_{2}$, in the following we assume that the SM particles are massless because of the DM particles masses are always much larger than the SM particles. Then the decay width of Eq.16 is given by
\begin{eqnarray}
\Gamma =\frac{\lambda^{4}m_{\chi _{1}}^{5}}{96\pi^{3}M_{\rm pl}^{4}}[1-8y+8y^{3}-y^{4}-12y^{2}{\rm ln}y],
\end{eqnarray}
where $y=\frac{m_{\chi_{2}}}{m_{\chi_{1}}}$, $m_{\chi _{1}}$ is the mass of DM $\chi_{1}$ and $m_{\chi _{2}}$ is the mass of DM $\chi _{2}$. 
We also assume that the coupling $\lambda \sim \mathcal{O}(1)$. 
To ensure the possibility of the decay process $\chi_{1}\rightarrow \chi_{2}\varphi_{1}\varphi_{2}$ We require that the DM particle $\chi_{2}$ should light than the DM particle $\chi_{1}$ i.e. $y<1$. Then to have a solution for Eq.17, we should also require 
\begin{eqnarray}
\frac{ 96\pi^{3}M_{\rm pl}^{4}\Gamma}{{\lambda^{4}}m_{\chi_{1}}^{5}}\lesssim \mathcal{O}(1).
\end{eqnarray}
This condition is presented in the right panel of FIG.6 (solid red line). In the right panel of FIG.6 we also give the limits from AMS-02 and IceCube \cite{ic2015}, we can see that the decay width $\Gamma \gtrsim 10^{-55}\rm TeV$ has been excluded. We also calculate the decay width $\Gamma$ for $y\rightarrow 1$ and $y=0.5$, ressult see the dot red line and solid blue line in right panel of FIG.6. Thus the theory allowed region is between the red solid and red dot line, and for $y=0.5$ we find that the the allowed DM mass range is $m_{\chi}\lesssim 3300\rm TeV$.

\begin{figure}
\includegraphics[width=80mm,angle=0]{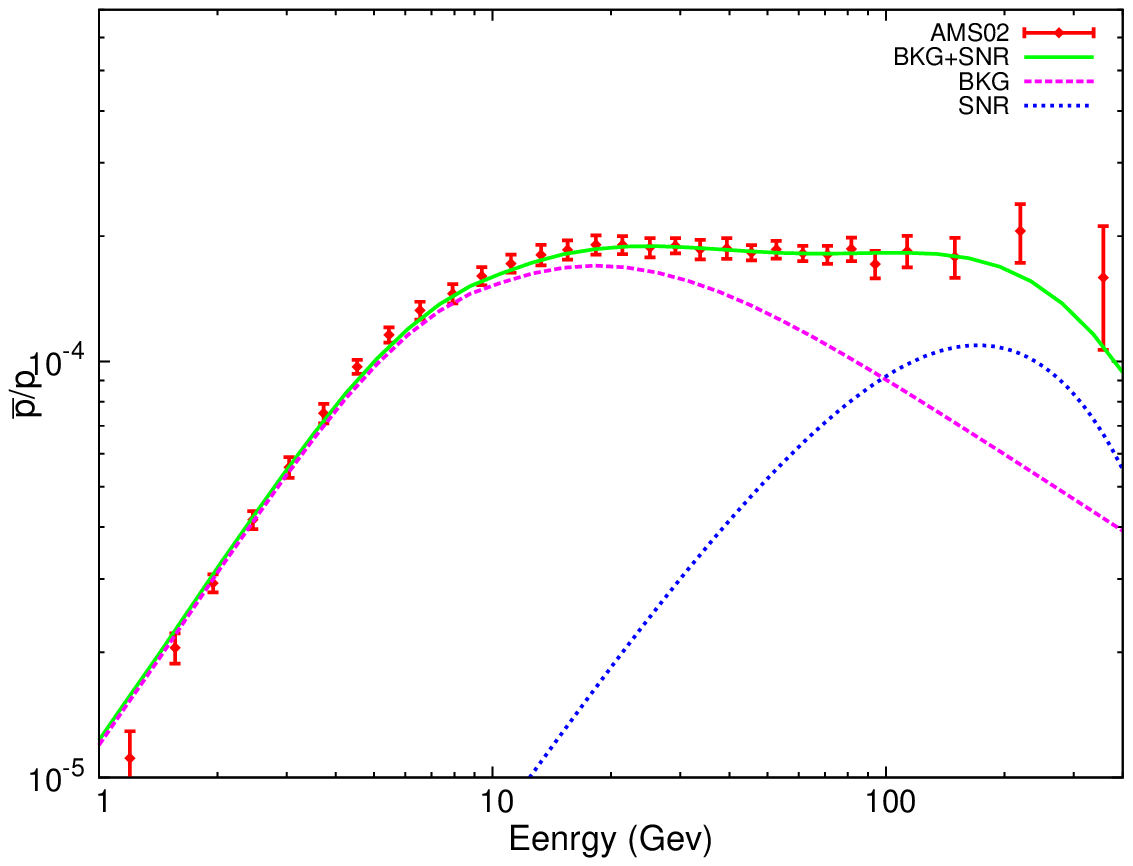}
\includegraphics[width=80mm,angle=0]{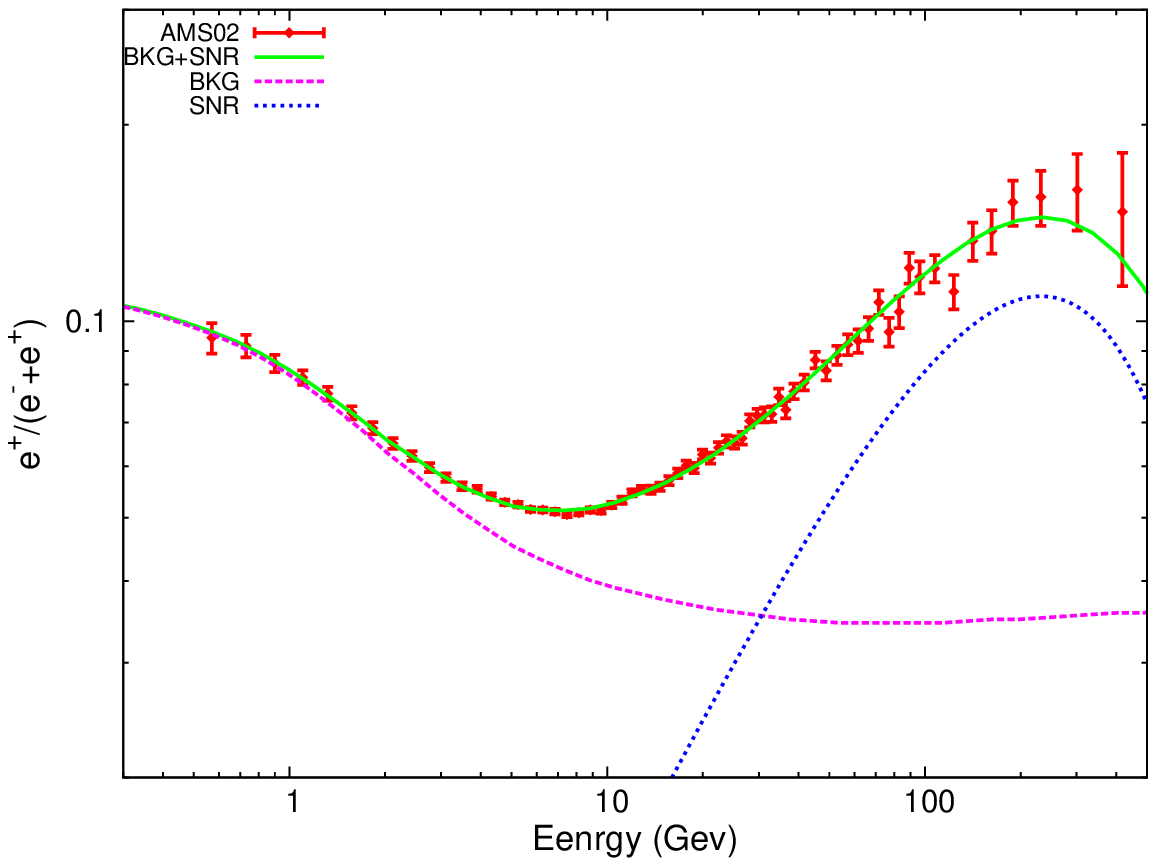}
\caption{Fit to the AMS-02 antiproton ratio and positron fraction data with BKG component and SNR component.}
\label{FIG.1}
\end{figure}

\begin{figure}
\includegraphics[width=80mm,angle=0]{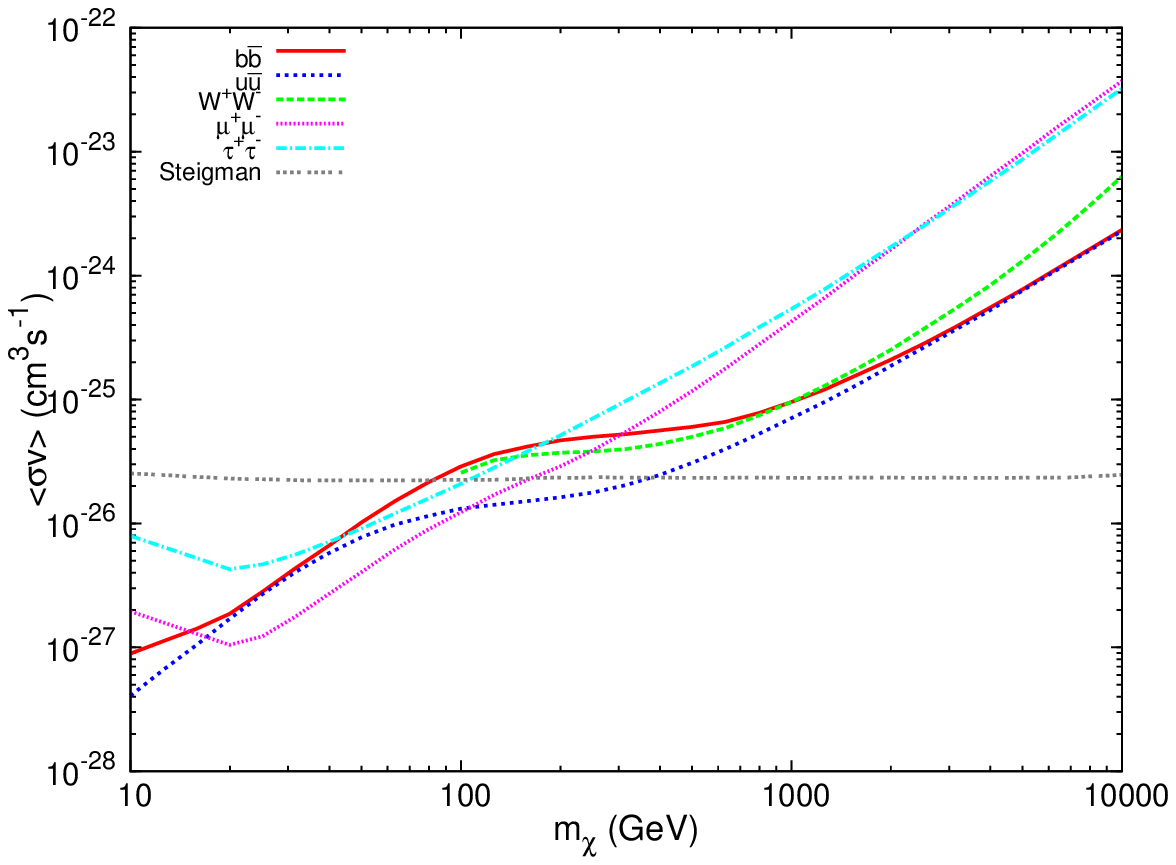}
\includegraphics[width=80mm,angle=0]{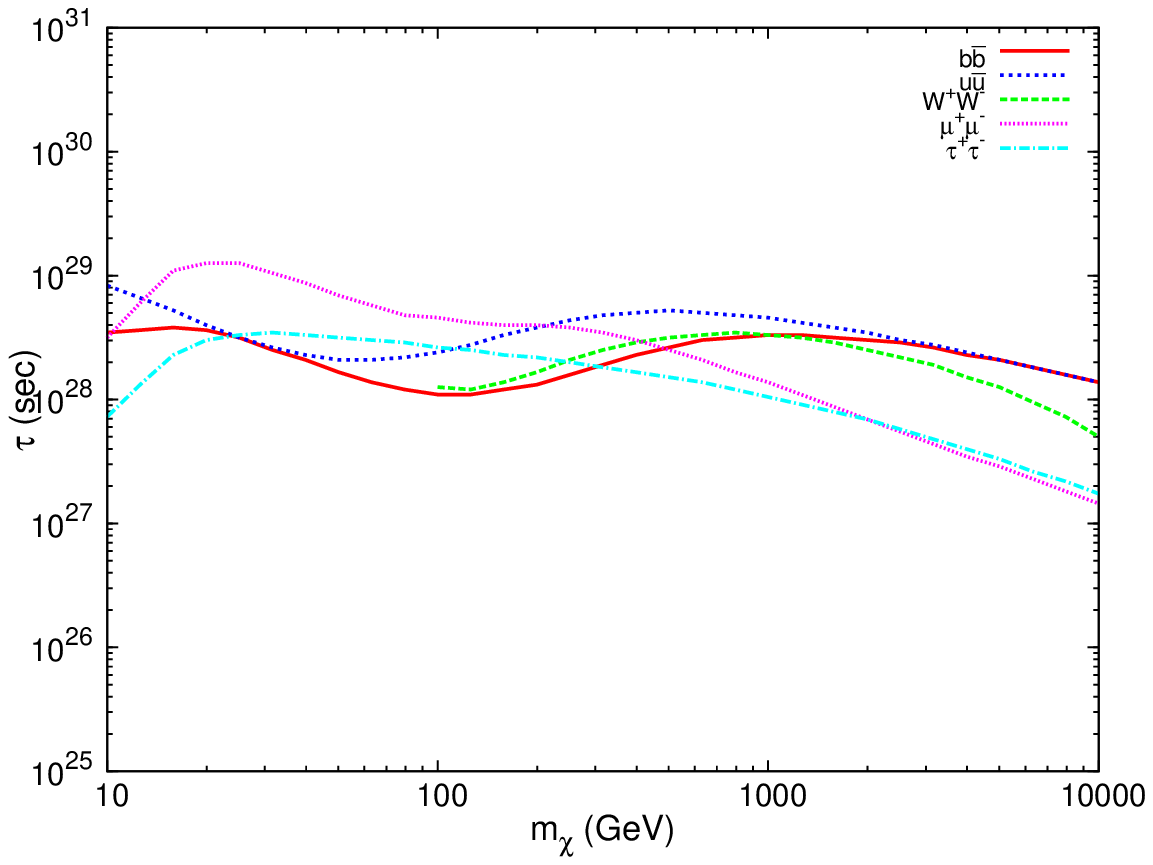}
\caption{
Left panel: limits on the DM annihilation cross section at 95$\%$ C.L. derived from the AMS-02 data. 
The gray curve in this and subsequent figures corresponds to the thermal relic cross section adopted from Steigman et al. \cite{Steigman2012}.
Right panel: limits on the DM annihilation cross section at 95$\%$ C.L. derived from the AMS-02 data.
The positron fraction is used to calculate the limits for the final states $\mu ^{+}\mu ^{-}$ and $\tau ^{+}\tau ^{-}$ while antiproton ratio is used to calculate the limits for the final states $b\bar{b}$, $u\bar{u}$ and $W^{+}W^{-}$. 
}
\label{FIG.2}
\end{figure}

\begin{figure}
\includegraphics[width=80mm,angle=0]{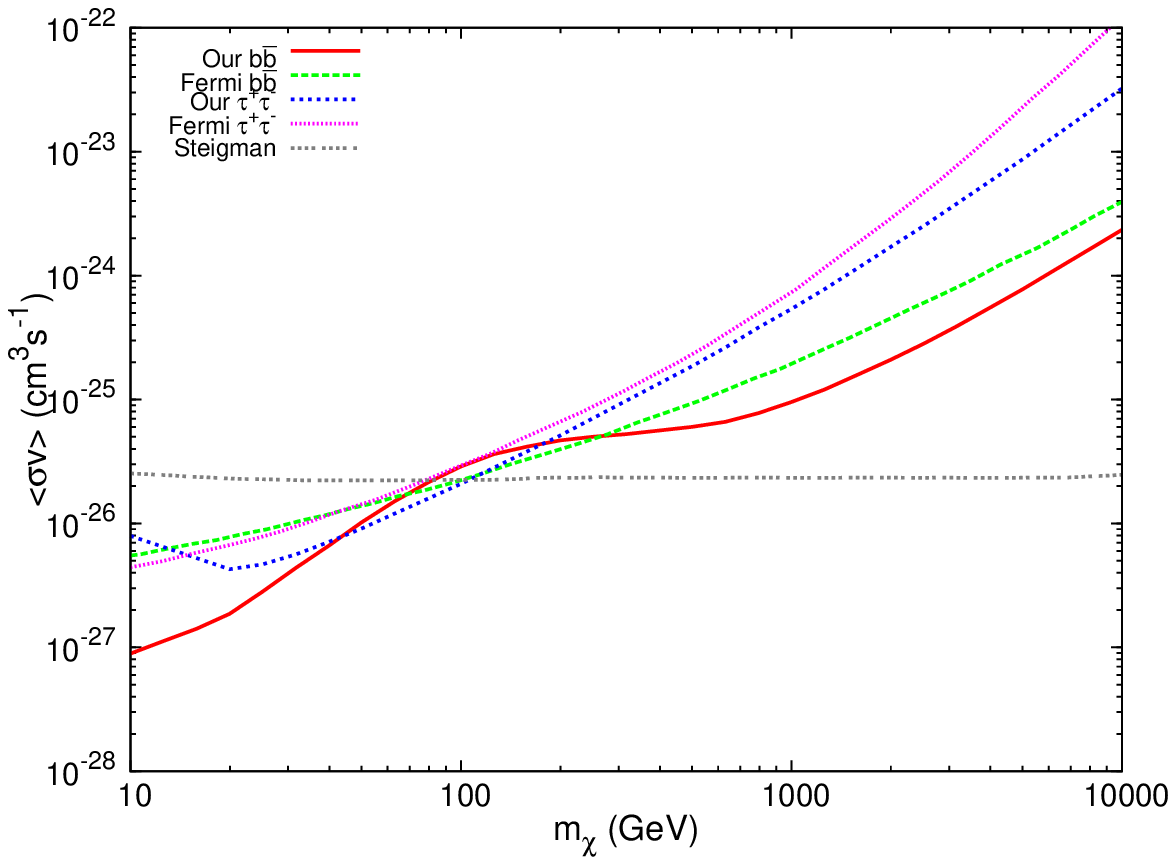}
\includegraphics[width=80mm,angle=0]{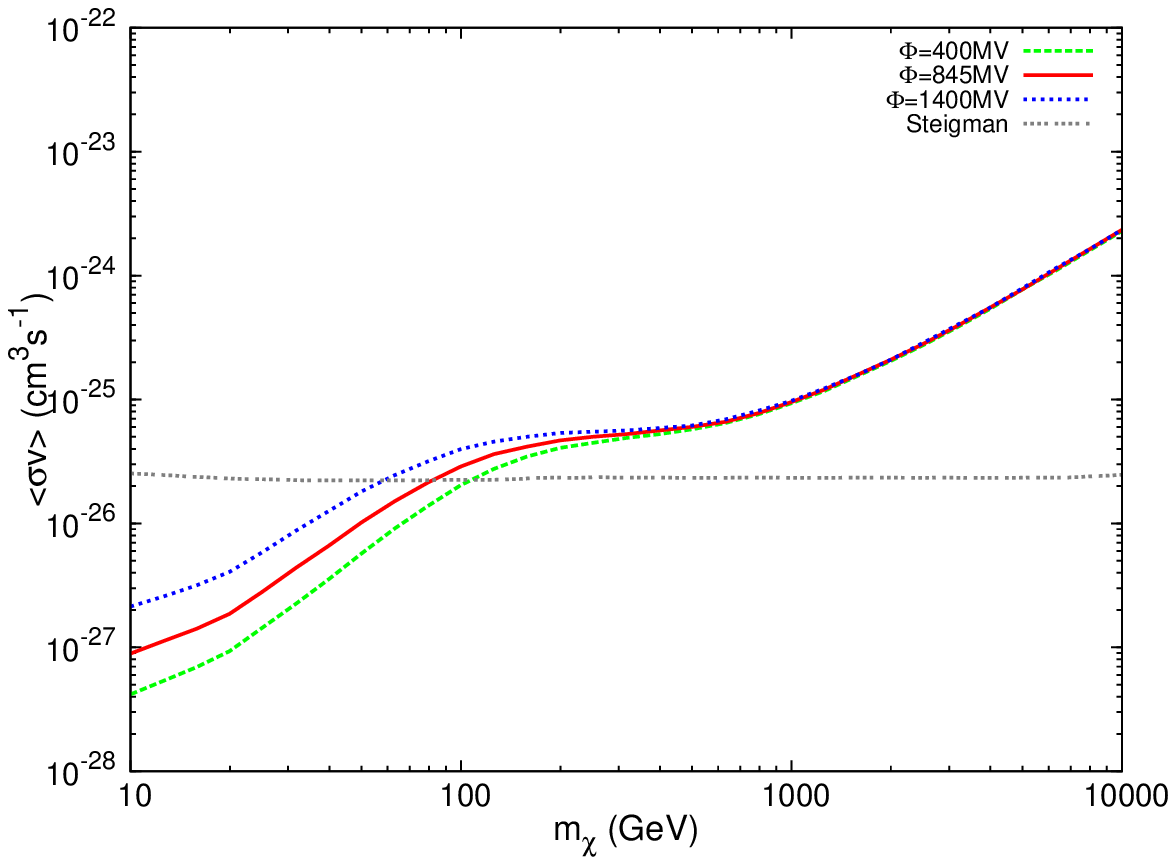}
\caption{
Left panel: compairson of our results with the upper limits from the Fermi-LAT observations of dwarf galaxies.
Right panel: the effect of the solar modulation parameter $\phi $ on the limit results.
}
\label{FIG.3}
\end{figure}

\begin{figure}
\includegraphics[width=80mm,angle=0]{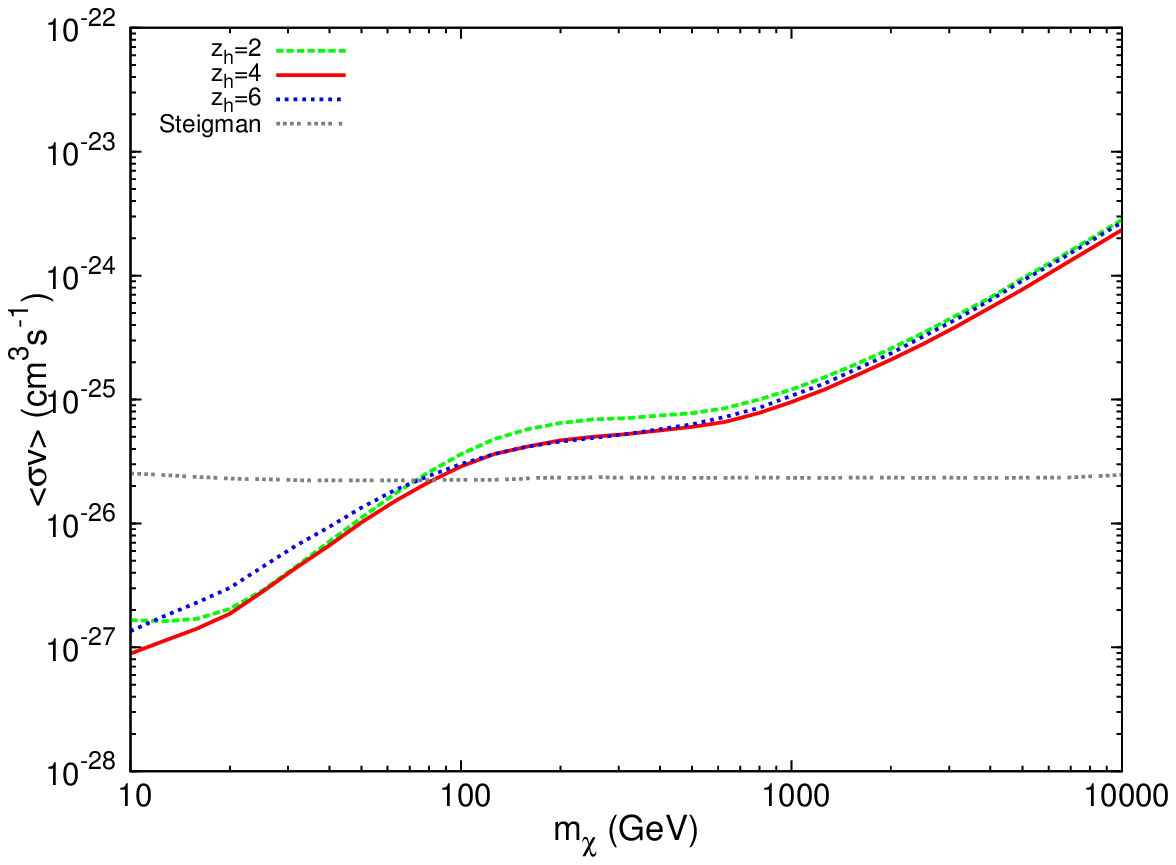}
\includegraphics[width=80mm,angle=0]{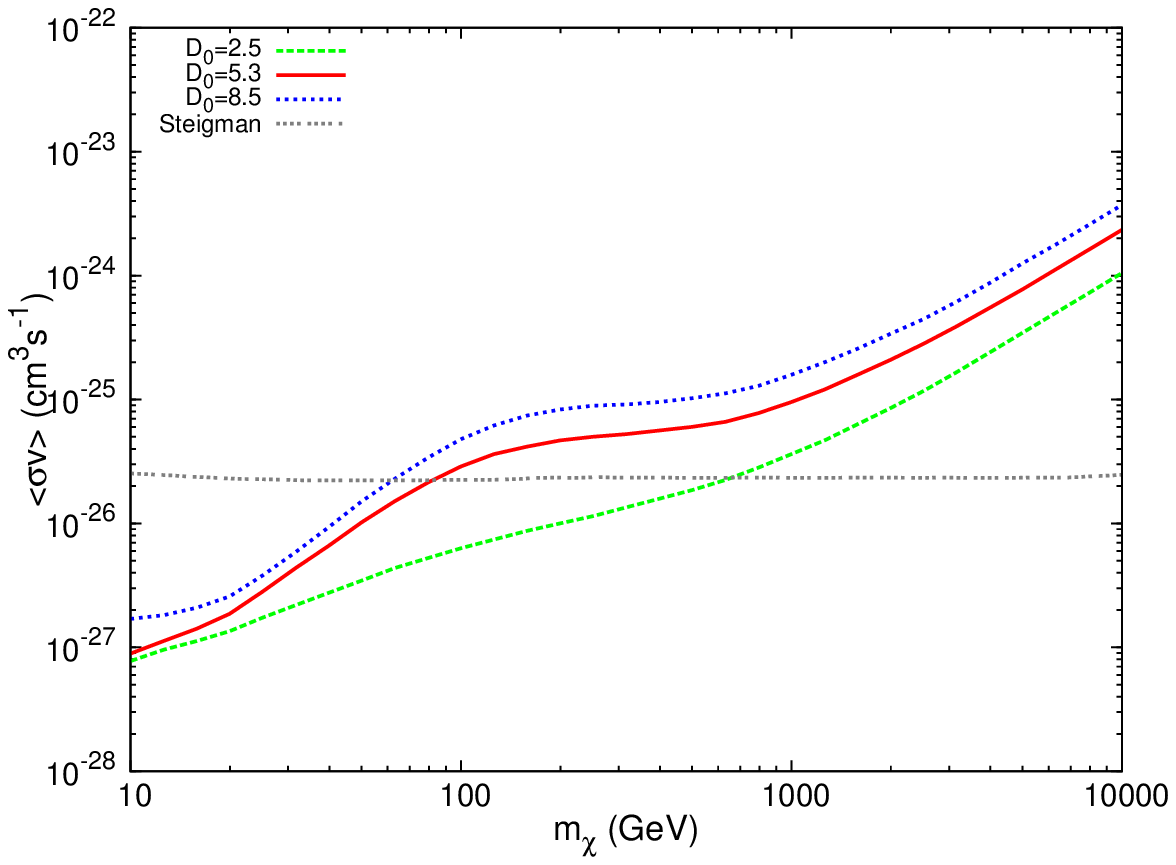}
\includegraphics[width=80mm,angle=0]{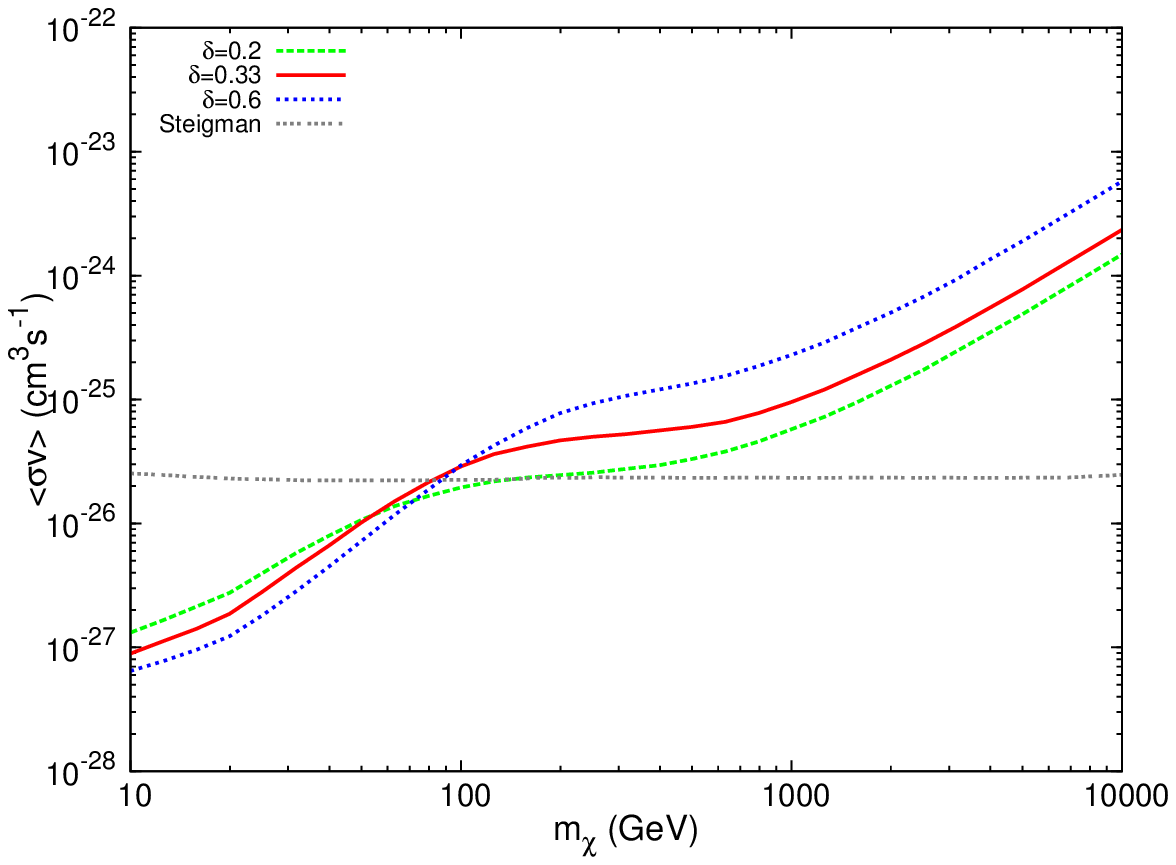}
\includegraphics[width=80mm,angle=0]{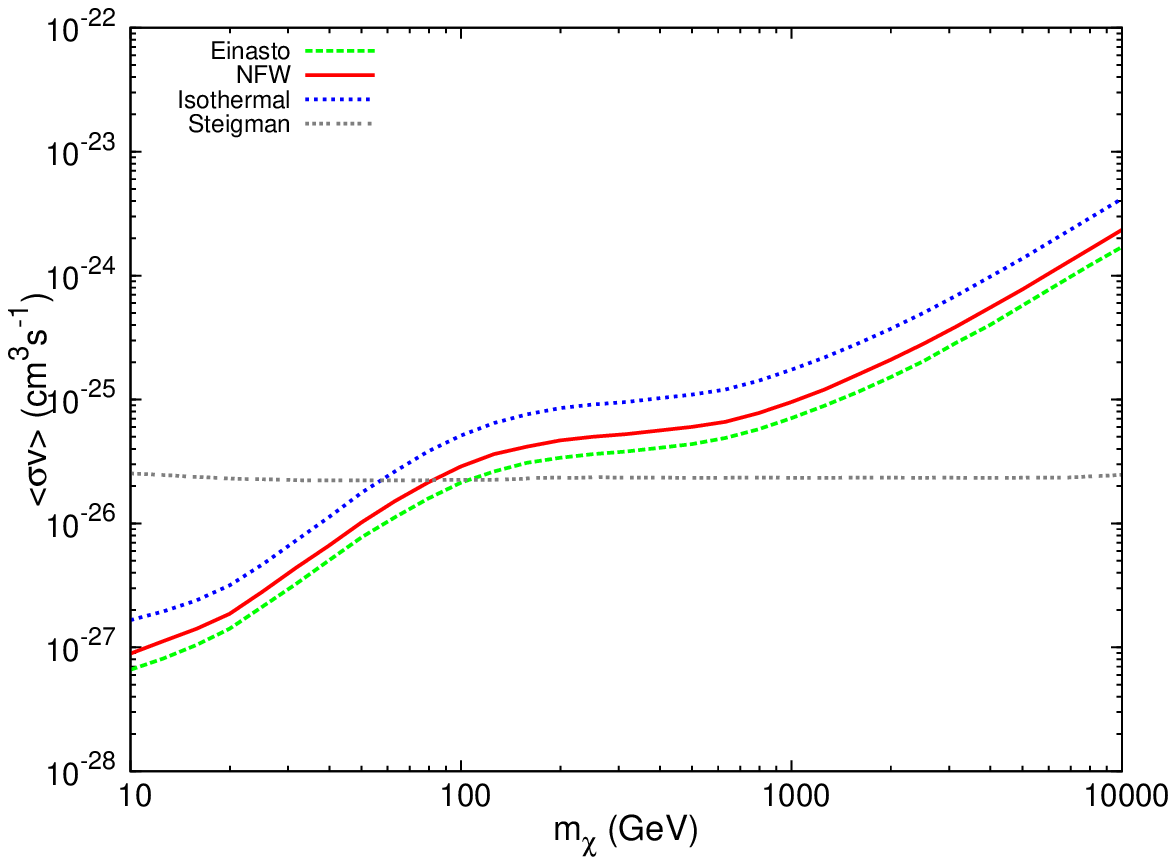}
\caption{
The effect of the propagation parameter and DM distribution profile.
}
\label{FIG.4}
\end{figure}

\begin{figure}
\includegraphics[width=80mm,angle=0]{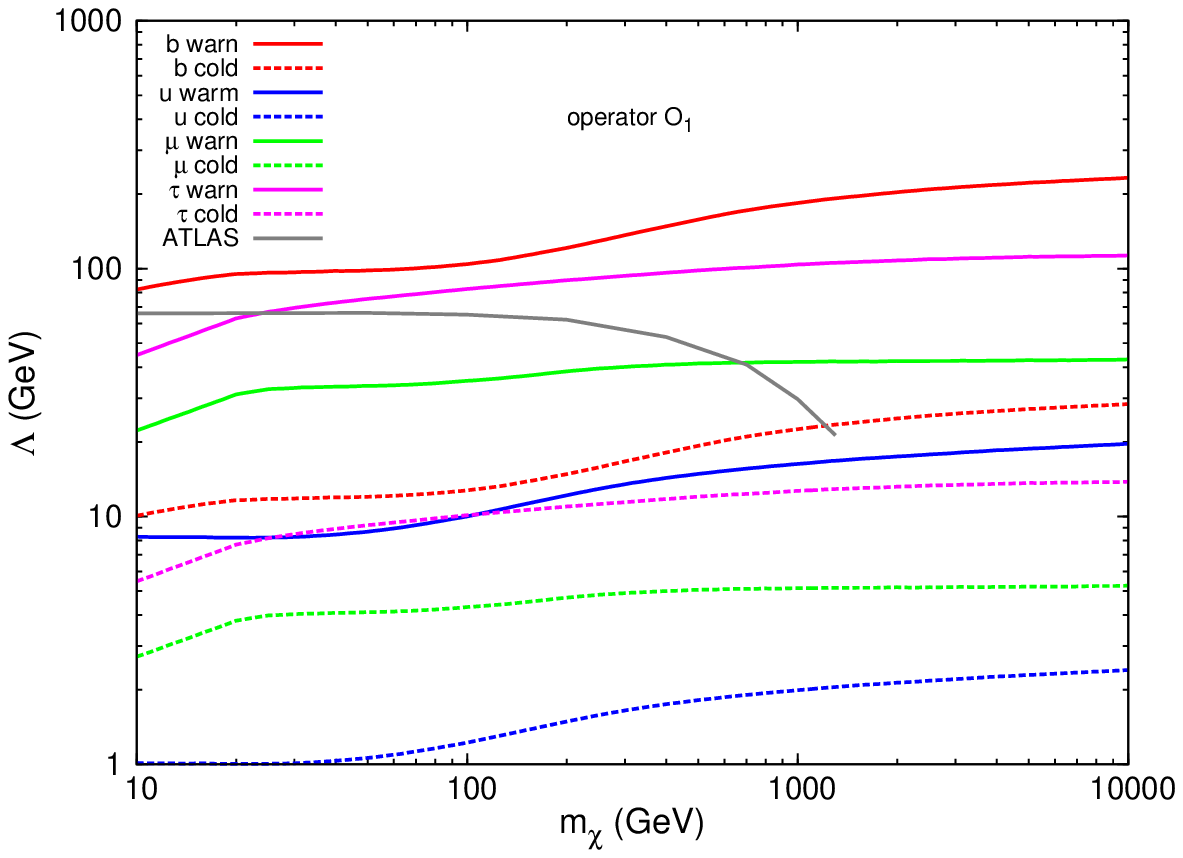}
\includegraphics[width=80mm,angle=0]{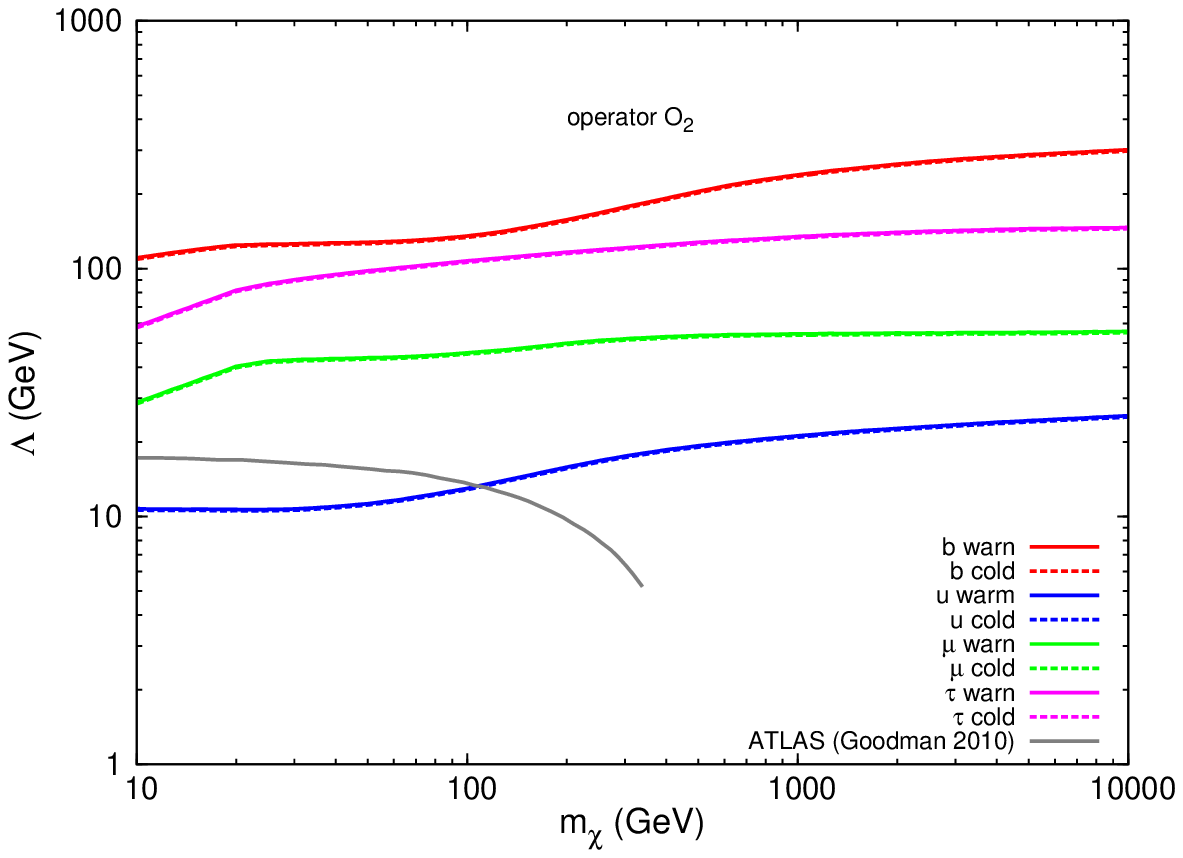}\\
\includegraphics[width=80mm,angle=0]{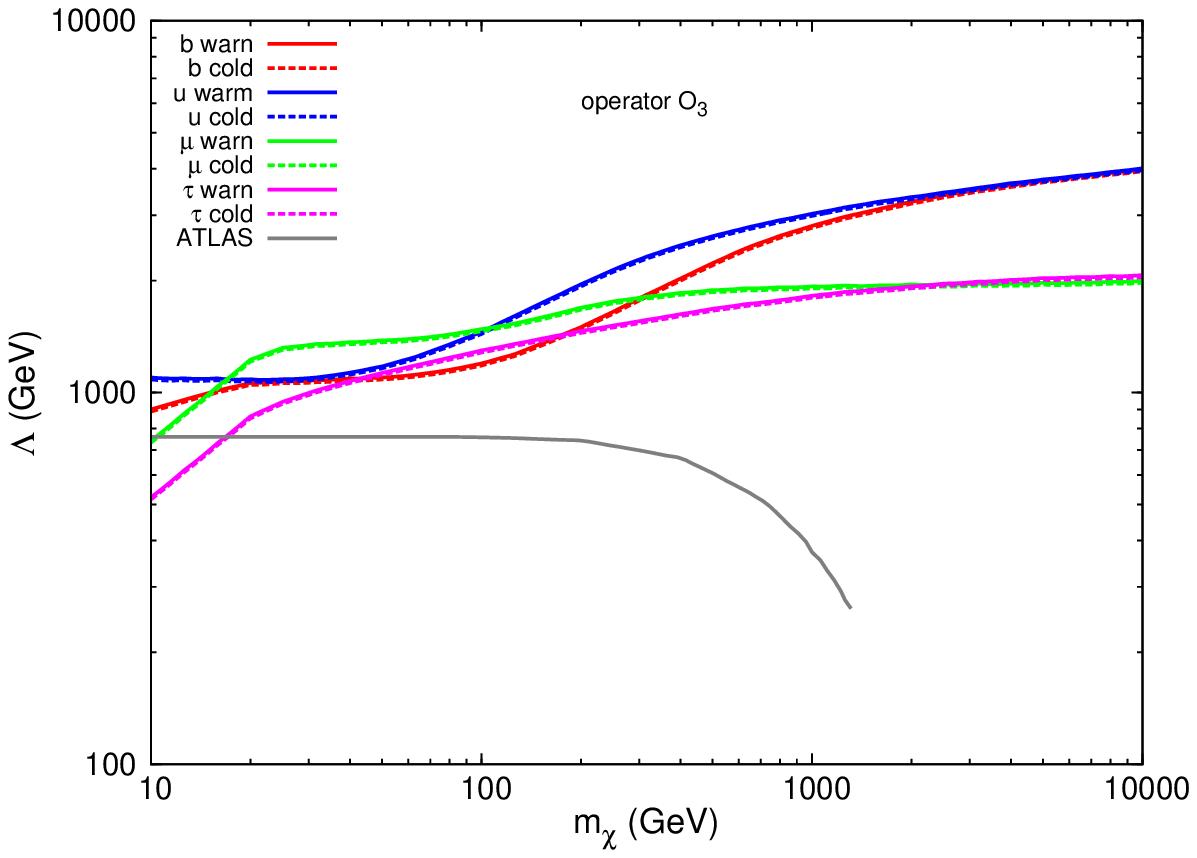}
\includegraphics[width=80mm,angle=0]{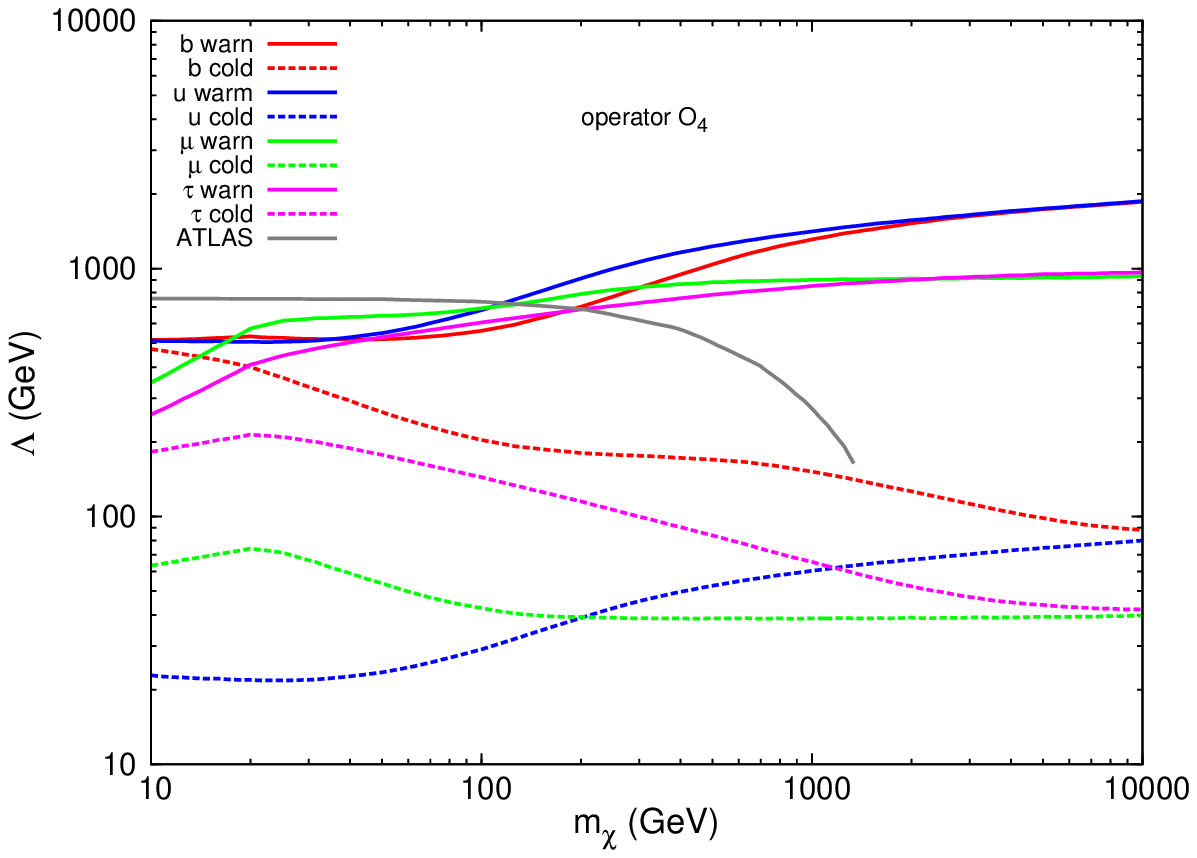}
\caption{
Limits on $\Lambda $ as a function of the DM particle mass $m_{\chi }$ at 95$\%$ C.L.,
for various operators as described in the texts. For each oprrator
final state of b and u quarks, $\mu $ and $\tau$ leptons are calculated, 
and for each final state we consider the case of warm DM and cold DM respectively,
the ATLAS results are from \cite{ATLAS1,goodman2010,ATLAS2,ATLAS3}.
}
\label{FIG.5}
\end{figure}

\begin{figure}
\includegraphics[width=80mm,angle=0]{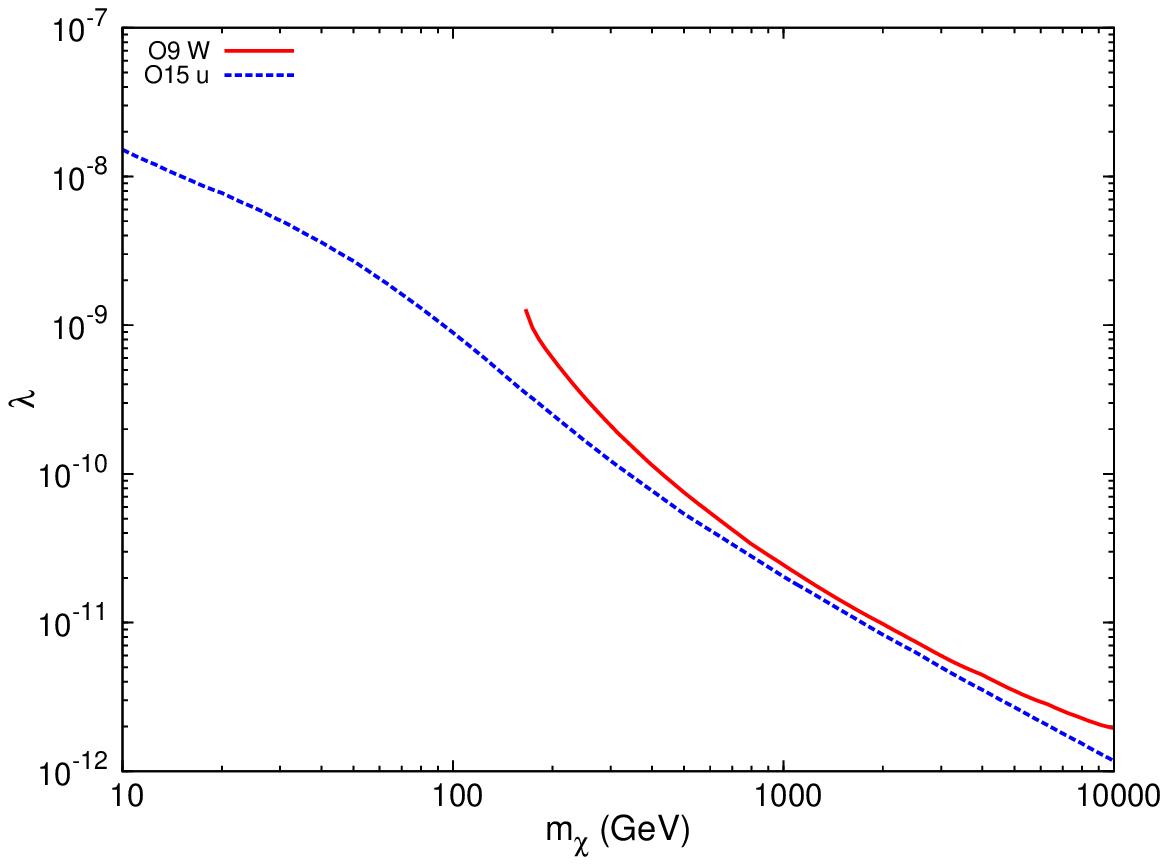}
\includegraphics[width=80mm,angle=0]{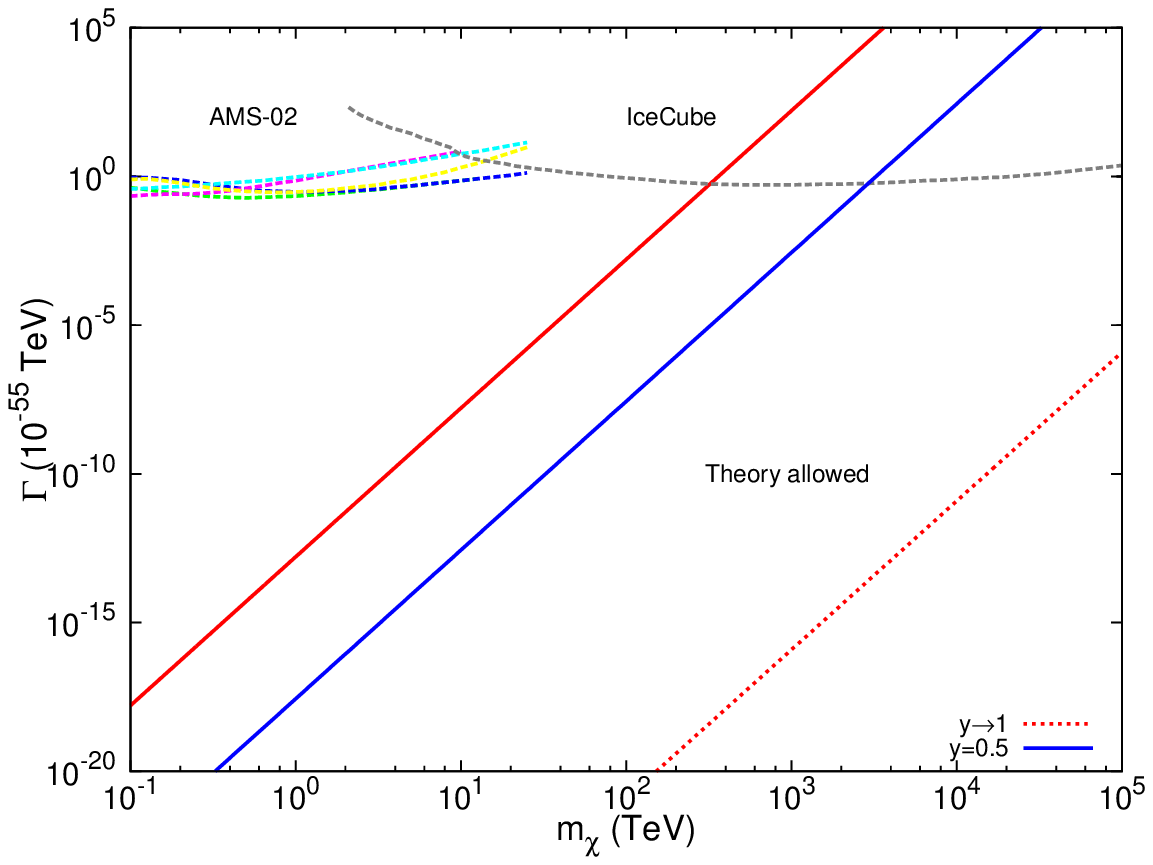}
\caption{
Left panel: limits on $\lambda $ as a function of the DM particle mass $m_{\chi }$ at 95$\%$ C.L..
Right panel: limits on the decay width $\Gamma $ of interaction $\mathcal{H}_{\rm VA}$ (Eq.16) as a function of the DM particle mass $m_{\chi}$, the region between the red solid line and red dot line is allowed by the theory and the region above the dashed line is excluded by AMS-02 and IceCube \cite{ic2015}.
}
\label{FIG.6}
\end{figure}


\section{Summary}
In this work, we have derived limits on the dark matter annihilation cross section and lifetime using measurements of the AMS-02 antiproton ratio and positron fraction data.
In deriving the limits, we consider the scenario of secondary particles accelerated in SNR which can explain the AMS-02 positron fraction and antiproton data at the same time, and then we parameterize the contribution of the SNR and calculate the BKG ratio of positron and antipropton by using GALPROP, then we add the SNR component to the BKG component as the total ratio at TOA. We use the likelihood ratio test to determine the significance of a possible DM contribution to the antiproton ratio and positron fraction measured by AMS-02. Upper limits at the $95\%$ $\rm C.L.$ on the DM annihilation or decay rate are derived by increasing the signal normalization from its best-fit value of background model, in this way we get the exclusion regions of DM parameters, including the annihilation cross section and lifetime for the final states $b\bar{b}$, $u\bar{u}$, $W^{+}W^{-}$ $\mu^{+}\mu^{-}$ and $\tau^{+}\tau^{-}$ as a function of $m_\chi$, respectively. Specifically, the positron fraction is used to calculate the constraints on the annihilation cross section and lifetime for the final states $\mu ^{+}\mu ^{-}$ and $\tau ^{+}\tau ^{-}$ while antiproton ratio is used to calculate the constraints on the annihilation cross section and lifetime for the final states $b\bar{b}$, $u\bar{u}$ and $W^{+}W^{-}$.
We find that our limits are stronger than the limits given by Ackermann et al. \cite{Ackermann} which derived from the Fermi-LAT gamma-ray Pass 8 data on the dwarf spheroidal satellite galaxies.

We also consider the uncertainty in our results and find that the propagation parameters contribute most uncertainty at large DM mass (above $\sim 100$ GeV) while the most uncertainty at low DM mass is contributed from solar modulation. As a result, the uncertainty of the limits on the DM parameters is about $(20-30)\%$ in the whole DM mass range if we take into account the contributions of propagation parameters and solar modulation, this value raises to $(40-50)\%$ if the contributions of DM distribution profile is taken into consideration.

Using this results, we also put limits on the suppression scale $\Lambda $ of effective field theory as a function of the DM particle mass $m_{\chi}$ for DM annihilation, and we also propose an effective interaction operator which may account for the stability of DM.

{\it Acknowledgments.} This work is supported in part by the National Natural Science Foundation of China (under Grants No. 11275097 and No. 11475085).

\end{document}